\documentclass[11pt,twoside]{article}
\usepackage{amssymb,color}%,pdfsync}
\newtheorem{thm}{Theorem}
\newtheorem{prop}[thm]{Proposition}
\newtheorem{corollary}[thm]{Corollary}
\newtheorem{lemma}[thm]{Lemma}
\newcommand{\N}{\mathbb N}
\newcommand{\R}{\mathbb R}
\newcommand{\ix}[1]{\int_{\R^d}#1\; dx}
\newcommand{\eqn}[1]{(\ref {#1})}
\newcommand{\nrm}[2]{\|{#1}\|_{L^{#2}(\R^d)}}
\newcommand{\be}[1]{\begin{equation}\label{#1}}
\newcommand{\ee}{\end{equation}}
\newcommand{\proof}{\noindent{\sl Proof.\/}\ }
\newcommand{\finprf}{\unskip\null\hfill$\square$\vskip 0.3cm}

\newcommand{\tr}{{\rm T\kern 0.2pt r}}
\newcommand{\seq}[1]{(#1)_{i\in\N^*}}
\newcommand{\scal}[2]{(#1,#2)_{L^2(\R^d)}}
\newcommand{\bnu}{{\mbox{\boldmath $\nu$}}}
\newcommand{\bpsi}{{\mbox{\boldmath $\psi$}}}
\newcommand{\Sp}{\R_+^{\N^*}\times (L^2(\R^d))^{\N^*}}
\newcommand{\Spp}{S\times (L^2(\R^d))^{\N^*}}
\newcommand{\F}{{\cal F}}
\newcommand{\email}[1]{{\small E-mail: {\textsf {#1}}}}
\newcommand{\http}[1]{{\small Internet: {\textsf {#1}}}}
\newcommand{\keywords}{\medskip\noindent{\bf Keywords. }}
\newcommand{\AMS}{\noindent{\small\bf AMS MSC (2000). }}

\newcommand{\CLT}{C_{\rm LT}(\gamma)}
\newcommand{\CLTd}{{\mathcal C}(\gamma)}
\newcommand{\CLTc}{C_{\rm LT}^{(1)}(\gamma)}

\newcommand{\CLTcd}{{\mathcal C}^{(1)}(\gamma)}
\newcommand{\CGN}{C_{\rm GN}(\gamma)}
\newcommand{\CGNd}{C_{\rm GN}^{\kern 1.5pt *}(\gamma)}
\newcommand{\CMain}{\CLTd}

\newcommand{\CF}{{\mathcal C}^{(1)}_F}

\definecolor{darkgreen}{rgb}{0,0.4,0}
\definecolor{darkred}{rgb}{0.8,0,0}
\definecolor{darkblue}{rgb}{0,0,0.7}
\begin{document}
\title{\sl Lieb-Thirring type inequalities and\\ Gagliardo-Nirenberg inequalities for systems}
\newcommand\runninghead[2]{\pagestyle{myheadings}\markboth{{\footnotesize\it\hfill{#1\quad}}}{{\footnotesize\it{#2\hfill\quad}}}}\headsep=40pt
\runninghead{J. Dolbeault, P. Felmer, M. Loss \& E. Paturel\qquad ---\qquad June 10, 2005}{\today\qquad ---\qquad Lieb-Thirring and Gagliardo-Nirenberg inequalities}
%%%%%%%%%%%%%%%%%%%%%%%%%%%%%%%%%%%%%%%%%%%%%%%%%%%%%%%%%%%%%%%%%%%%%%
\author{J. Dolbeault\footnote{Ceremade (UMR CNRS no. 7534), Universit\'e Paris Dauphine, Place de Lattre de Tassigny, 75775 Paris C\'edex~16, France. Tel: (33) 1 44 05 46 78, Fax: (33) 1 44 05 45 99. \email{dolbeaul@ceremade.dauphine.fr}, \http{http://www.ceremade.dauphine.fr/$\sim$dolbeaul/}},
P. Felmer\footnote{Departamento de Ingenier\'{\i}a Matem\'atica and Centro de Modelamiento Matem\'atico, UMR CNRS - UChile 2071, Universidad de Chile, Casilla 170 Correo 3, Santiago, Chile. \email{pfelmer@dim.uchile.cl}},
M. Loss\footnote{School of Mathematics, Georgia Institute of Technology, Atlanta, GA 30332, USA. Tel: (1) 404 894 271, Fax: (1) 404 894 4409. \email{loss@math.gatech.edu}, \http{http://www.math.gatech.edu/$\sim$loss/}},
E. Paturel\footnote{Laboratoire de Math\'ematiques Jean Leray - Universit\'e de Nantes, 2, rue de la Houssini\`ere - 44322 Nantes Cedex 3, France. Tel: (33) 2 51 12 59 57, Fax: (33) 2 51 12 59 12. \email{Eric.Paturel@math.univ-nantes.fr}, \http{http://www.math.sciences.univ-nantes.fr/$\sim$paturel/}}}
\date{June 10, 2005}
\maketitle
\thispagestyle{empty}
%%%%%%%%%%%%%%%%%%%%%%%%%%%%%%%%%%%%%%%%%%%%%%%%%%%%%%%%%%%%%%%%%%%%%%
\abstract{\sl This paper is devoted to inequalities of Lieb-Thirring type. Let $V$ be a nonnegative potential such that the corresponding Schr\"odinger operator has an unbounded sequence of eigenvalues $(\lambda_i(V))_{i\in\N^*}$. We prove that there exists a positive constant ${{\mathcal C}(\gamma)}$, such that, if~$\gamma>d/2$, then
$$
\sum_{i\in\N^*}\left[\lambda_i(V)\right]^{-\gamma}\leq {{\mathcal C}(\gamma)}\int_{\R^d}{V^{\frac d2-\gamma}}\;dx
\eqno(*)$$
and determine the optimal value of ${{\mathcal C}(\gamma)}$. Such an inequality is interesting for studying the stability of mixed states with occupation numbers. 

We show how the infimum of $\lambda_1(V)^\gamma\kern-1.5pt\cdot\kern-1.5pt\int_{\R^d}{V^{\frac d2-\gamma}}\,dx$ on all possible potentials $V$, which is a lower bound for $[{{\mathcal C}(\gamma)}]^{-1}$, corresponds to the optimal constant of a subfamily of Gagliardo-Nirenberg inequalities. This explains how $(*)$ is related to the usual Lieb-Thirring inequality and why all Lieb-Thirring type inequalities can be seen as generalizations of the Gagliardo-Nirenberg inequalities for systems of functions with occupation numbers taken into account. 

We also state a more general inequality of Lieb-Thirring type
$$
\sum_{i\in\N^*}F(\lambda_i(V))=\tr \left[ F\left(-\Delta +V\right)\right]\leq\int_{\R^d}G(V(x))\,dx
\eqno(**)$$
where $F$ and $G$ are appropriately related. As a special case corresponding to $F(s)=e^{-s}$, $(**)$ is equivalent to an optimal euclidean logarithmic Sobolev inequality
$$
\int_{\R^d}{\rho\,\log\rho}\;dx+\frac d2\,\log(4\pi)\int_{\R^d}{\rho}\;dx\leq \sum_{i\in\N^*}\nu_i\,\log\nu_i+\sum_{i\in\N^*}\nu_i\int_{\R^d}{|\nabla\psi_i|^2}\;dx
$$
where $\rho=\sum_{i\in\N^*}\nu_i\,|\psi_i|^2$, $\seq{\nu_i}$ is any nonnegative sequence of occupation numbers and $\seq{\psi_i}$ is any sequence of orthonormal $L^2(\R^d)$ functions.
}\smallskip

\medskip

\AMS{\scriptsize Primary: 26D15, 52A40, 35P20; Secondary: 35J10, 47A75, 49R50, 81Q10}
\section{Introduction}\label{Sec:Intro}
Lieb-Thirring type inequalities are well known in the context of the stability of matter in quantum mechanics. Let $\hbar=h/2\pi>0$ and $m>0$ be respectively Planck's constant and the mass constant. Given a smooth bounded nonpositive potential $V$ on $\R^d$, if we denote by 
\[
\lambda_1(V)<\lambda_2(V)\leq\lambda_3(V)\leq\ldots\,\lambda_N(V)<0
\]
the finite sequence of all negative eigenvalues of the Schr\"odinger operator 
\[
H_V=-\frac{\hbar^2}{2m} \Delta+V\;,
\]
then it is possible to bound the sum $\sum_{i=1}^N|\lambda_i(V)|^\gamma$ in terms of $\|V\|_{L^{\gamma+d/2}(\R^d)}$ whatever $N$ is. The inequality
\be{Ineq:LT}
\sum_{i=1}^N|\lambda_i(V)|^\gamma\leq \CLT\ix{|V|^{\gamma+\frac d2}}
\ee
is known as the {\sl Lieb-Thirring inequality.\/} Here we denote by $\CLT$ the smallest possible positive constant which is independent of $V$. For $\gamma=1$, the sum $\sum_{i=1}^N|\lambda_i(V)|$ is the {\sl complete ionization energy,\/} which is the physically relevant quantity for studying the stability of matter. Considerable efforts have been made to understand further the Lieb-Thirring inequality and in particular to find the optimal value of $\CLT$. Up to now a few facts about the sharp constant in the Lieb-Thirring inequality are known. It was proved in \cite{MR598768} for $d=1$ and later generalized to arbitrary $d$ in \cite{Laptev-Weidl00} that for $\gamma \ge 3/2$ the sharp constant is given by the semiclassical constant, {\sl i.e.\/} the constant corresponding to the limit problem when letting $\hbar\to 0$, after an appropriate scaling. Among many other open problems, the {\sl Lieb-Thirring conjecture\/} asserts that in $d=1$
\[\label{Eqn:Lieb-ThirringConjecture}
\CLT=\CLTc:=\inf_{\begin{array}{c} V\in{\cal D}(\R)\\ V\leq 0\end{array}} \frac{|\lambda_1(V)|^\gamma}{\int_{\R}|V|^{\gamma+\frac 12}\;dx}\;.
\]
This has been worked out for the case $\gamma= 1/2$ in \cite{MR2000c:81062}. Also see \cite{MR1079775,MR1061661,Lieb-Thirring76,Lieb-Thirring,Hundertmark-Laptev-Weidl00,Benguria-Loss,MR2001i:00001} for further results on \eqn{Ineq:LT}.

\medskip What we study in this note is a somewhat different problem, where $V$ is a nonnegative, unbounded potential on $\R^d$, such that the eigenvalues of $H_V$ form a positive unbounded nondecreasing sequence $(\lambda_i(V))_{i\in\N^*}$. Our main result is the
%-----------------------------------------------------------------------------
\begin{thm}\label{Thm:Main} For any $\gamma>d/2$, $d\in\N^*$, and for any nonnegative $V\in C^\infty(\R^d)$ such that $V^{d/2-\gamma}\in L^1(\R^d)$,
\be{main}
\sum_{i\in\N^*}\left[\lambda_i(V)\right]^{-\gamma}\leq \CMain\ix{V^{\frac d2-\gamma}}\;.
\ee
The value of the sharp constant $\CMain$ is given by the Weyl asymptotics, {\rm i.e.\/} by its value in the semiclassical limit:
\[
\CMain=\left(\frac{m}{2 \pi \hbar^2}\right)^{d/2} \frac{\Gamma(\gamma -d/2)}{\Gamma(\gamma)}\;.
\]
\end{thm}
%-----------------------------------------------------------------------------
Although the result is quite simple and arises as an immediate consequence of the Golden and Thompson inequality, it is to our knowledge new. Some partial estimates have been obtained in \cite{MR1686426} in case of quadratic potentials. We refer to \cite{MR50:2996,MR49:8556,MR84k:58225,Federbush} for earlier results in this direction. The semiclassical formula stems from prescribing $h^d$ phase space volume to each bound state of the Schr\"odinger operator. Using this heuristics, we can estimate the series by
\[
\sum_{i\in\N^*}\left[\lambda_i(V)\right]^{-\gamma}\approx\frac{1}{h^d}\int_{\R^d\times\R^d} \left(\frac{p^2}{2m}+V(x)\right)^{\!\!-\gamma}dp\,dx\;,
\]
which is easily seen to yield the right hand side of (\ref{main}). If one considers $h$ as a parameter, then as $h$ gets small the two sides in the above relation are asymptotically the same. A rigorous proof of this fact relies on the Weyl asymptotics, that we will establish later in this paper for a specific potential $V$.

We may notice that all physical constants can be adimensionalized. A simple scaling indeed shows that for any $i\in\N^*$, 
\[
\lambda_i^{\hbar,\,m}(V)=\lambda_i^{1,\,1/2}\left(V\left({\scriptstyle \frac\hbar{\sqrt{2m}}}\cdot\right)\right)\;,
\]
so that in \eqn{main}, 
\[
\CMain_{|\hbar,\,m}=\left(\frac{2\,m}{\hbar^2}\right)^{d/2}\CMain_{|\hbar=1,\,m=1/2}\;,\quad \CMain_{|\hbar=1,\,m=1/2}=\frac{\Gamma\left(\gamma-\frac d2\right)}{(4\pi)^{d/2}\,\Gamma\left(\gamma\right)}\;.
\]
{}From now on, we assume for simplicity that $\hbar=1,m=1/2$. 

\medskip Theorem \ref{Thm:Main} is motivated by the study of the dynamical stability of mixed states with respect to minimizers of variational problems with temperature in quantum mechanics. Inequality \eqn{Ineq:LT} appears in the context of atomic and molecular physics, where it is natural to consider isolated systems for which the potential $V$ is asymptotically zero at infinity. Computing the full ionization energy is then a completely relevant question. Requiring that $V$ grows at infinity makes sense in a different context, {\sl e.g.,\/} in solid state physics, where the potential is not necessarily created by the system under consideration itself, but can be imposed by external devices (for instance a doping profile) or by a given electrostatic field (applied voltage). In that case, collective effects are fundamental and it is interesting to investigate how mixed states converge in a semi-classical limit to a classical system. At the kinetic level, the behavior of the classical system is now reasonably well understood. For instance one knows in which sense special stationary solutions are stable, see \cite{Braasch-Rein-Vukadinovic99,Caceres-Carrillo-Dolbeault03}. At the quantum level, many particle systems are not so well understood. A first attempt in this direction has been made in \cite{MR2003a:82066}, in a nonlinear case, but the result relies on a rather weak notion of stability and the exchange term is neglected. For a linear system, we will see in Section~\ref{Sec:Stability} that an appropriate functional for studying the stability of a mixed state, {\sl i.e.\/} a sequence $(\bnu,\bpsi)=(\nu_i,\psi_i)_{i\in\N^*}\in\Sp$ made of nonnegative ordered occupation numbers $\nu_i$ and wave functions $\psi_i$, is the {\sl free energy functional\/} 
\[
\F[\bnu,\bpsi]:= \sum_{i\in\N^*}\left[\beta(\nu_i)+\nu_i\,\int_{\R^d}\Big(|\nabla\psi_i|^2+V\,|\psi_i|^2\Big)\, dx\right]\,,
\]
where $\beta$ is a given convex function on $\R_+$. Under the constraints
\[
\scal{\psi_i}{\psi_j}=\delta_{ij}\quad\forall\; i,\, j\in\N^*\;,
\]
the functional $\F$ has a minimizer made of the sequence $\bar\bpsi=\seq{\bar\psi_i}$ of the eigenfunctions counted with multiplicity, and the sequence $\bar\bnu=\seq{\bar\nu_i}$ of occupation numbers given in terms of the eigenvalues by
\[\label{Eqn:OccupationMinimizer}
\bar\nu_i=(\beta')^{-1}(-\lambda_i(V))\;.
\]
However, such considerations are purely formal as long as we dont prove that $\F[\bnu,\bpsi]$ is finite at least for the formal minimizer $(\bnu,\bpsi)=(\bar\bnu,\bar\bpsi)$. Such a property is a condition on both $\beta$ and $V$. In case 
\be{Eqn:beta}
\beta(\nu)=\left\{\begin{array}{ll}
-(1-m)^{m-1}\,m^{-m}\,\nu^m\quad&\mbox{if}\;\nu\geq 0\;,\\
&\\
+\infty\quad&\mbox{if}\;\nu<0\;,
\end{array}\right.\qquad \mbox{and}\quad m\in (0,1)\;,
\ee
and with $\gamma=\frac m{1-m}$, for any $i\in\N^*$, we obtain
\begin{eqnarray*}
&&\bar\nu_i=\frac{m}{1-m}\left(\lambda_i(V)\right)^{\frac 1{m-1}}\,,\\
&&\beta(\bar\nu_i)+\bar\nu_i\,\lambda_i(V)=\left(\beta\circ (\beta')^{-1}\right)(-\lambda_i(V))+(\beta')^{-1}(-\lambda_i(V))\,\lambda_i(V)=-\left[\lambda_i(V)\right]^{-\gamma}.
\end{eqnarray*}
The free energy is well defined at least for the optimal mixed state if the series $\sum_{i\in\N^*}[\beta(\bar\nu_i)+\bar\nu_i\,\lambda_i(V)]$ converges, which amounts to require that $\sum_{i\in\N^*}\left[\lambda_i(V)\right]^{-\gamma}$ is finite. A sufficient condition is therefore given by Theorem \ref{Thm:Main}.

\bigskip Section \ref{Sec:ProofMain} is devoted to a proof of Theorem \ref{Thm:Main} based on the inequality of Golden and Thomson (Theorem \ref{Thm:Symanzik}). We also state a more general result in Theorem \ref{Thm:Main-Gen}. The notion of dynamic stability will be explained in Section~\ref{Sec:Stability} and illustrated by several examples. In Section \ref{Sec:Stability}, we will come back to the constant $\CLTc$ which appears in the Lieb-Thirring conjecture and prove that it is related to the best constant in some special Gagliardo-Nirenberg inequalities in the standard case corresponding to $V\leq 0$. Such a result is not new, but we also prove that a similar result holds in the case $V\geq 0$ (case of Theorem~\ref{Thm:Main}), which is apparently new. We also relate a limiting case to the euclidean logarithmic Sobolev inequality. In Section~\ref{Sec:Gagliardo-Nirenberg-Systems} we show in which sense Lieb-Thirring type inequalities can be seen as generalizations of the Gagliardo-Nirenberg inequalities to systems. This extends to systems of orthonormal functions what has been observed in Sections \ref{Sec:Lieb-ThirringConjecture} and~\ref{Sec:DualCase}. We formulate optimal inequalities in an abstract framework and apply the result to the standard case (Corollary \ref{Cor:Interp1}), to the framework of Theorem \ref{Thm:Main} (Corollary \ref{Cor:Interp2}) and to a limiting case which provides an optimal inequality of logarithmic Sobolev type for systems. Optimal constants are expressed in terms of the optimal constants for Lieb-Thirring type inequalities.

%%%%%%%%%%%%%%%%%%%%%%%%%%%%%%%%%%%%%%%%%%%%%%%%%%%%%%%%%%%%%%%%%%%%%%
%%%%%%%%%%%%%%%%%%%%%%%%%%%%%%%%%%%%%%%%%%%%%%%%%%%%%%%%%%%%%%%%%%%%%%
\section{Proof of Theorem \ref{Thm:Main}}\label{Sec:ProofMain}

The proof of Theorem \ref{Thm:Main} is straightforward. It relies on a change of variables in the definition of the $\Gamma$~function and the following inequality due to Golden and Thompson \cite{MR32:7113, MR32:7110}. See \cite{MR84m:81066, MR95e:81130} for an introduction to such methods and a proof based on the Feynman-Kac formula. Here we adopt the presentation of \cite{Symanzik} as stated in \cite{MR84m:81066}, Theorem~9.2, p. 94.
%-----------------------------------------------------------------------------
\begin{thm}\label{Thm:Symanzik}{\rm \cite{Symanzik, MR84m:81066}} Let $V$ be in $L^1_{\rm loc}(\R^d)$ and bounded from below. Assume moreover that $e^{-tV}$ is in $L^1(\R^d)$ for any $t>0$. Then 
\be{Ineq:Symanzik}
\tr \left(e^{-t\, (-\Delta +V)}\right)\leq (4\pi t)^{-\frac{d}{2}}\int_{\R^d}e^{-t\, V(x)}\,dx\;.
\ee\end{thm}
%-----------------------------------------------------------------------------
\proof For completeness, we give an elementary proof of this result. We do not claim originality here and we give this result only for the convenience of the reader.

Consider the Green function $G$ of the heat equation:
\[
G(x,t):=(4\pi t)^{-\frac{d}{2}}\,e^{-\frac{|x|^2}{4t}}\;.
\]
We will then write
\[
u(\cdot,t)=e^{t\,\Delta}\,f:=G(\cdot,t)*f
\]
if $u$ is a solution of $u_t=\Delta u$ with initial data $u(\cdot, t=0)=f$. By Trotter's formula, $e^{-t (-\Delta +V)}$ is obtained as the strong limit of 
\[
\left(e^{\frac tn\,\Delta }\,e^{-\frac tn\,V}\right)^n
\]
as $n$ goes to infinity. Then we compute the trace of this last quantity as
\begin{eqnarray*}
\int_{(\R^d)^n}dx\,dx_1\,dx_2\ldots\,dx_n\;G\left(\frac tn,x-x_1\right)\,e^{-\frac tn\,V(x_1)}\,G\left(\frac tn,x_1-x_2\right)\,e^{-\frac tn\,V(x_2)}\ldots&& \\ \ldots\; G\left(\frac tn,x_n-x\right)\,e^{-\frac tn\,V(x)}\;.&&
\end{eqnarray*}
With the notation $x=x_0=x_{n+1}$, we rewrite this as 
\[
\int_{(\R^d)^n}dx_0\,dx_1\,dx_2\ldots\,dx_n\;\prod_{j=0}^n G\left(\frac tn,x_j-x_{j+1}\right)\;e^{-\frac tn\,\sum_{k=0}^{n-1} V(x_k)}\;.
\]
Using the convexity of $x\mapsto e^{-x}$, we estimate the exponential term by:
\[
e^{-\frac tn\,\sum_{k=0}^{n-1} V(x_k)}\leq \frac 1n\,\sum_{k=0}^{n-1} e^{-t\,V(x_k)}\;.
\]
This amounts to
\begin{eqnarray*}
\tr\left(e^{\frac tn\,\Delta }\,e^{-\frac tn\,V}\right)^n&\leq& \frac 1n\,\sum_{k=0}^{n-1} \int_{(\R^d)^n}\kern -15pt dx_0\,dx_1\,dx_2\ldots\,dx_n\;\prod_{j=0}^n\! G\!\left(\frac tn,x_j-x_{j+1}\!\right)\,e^{-t\,V(x_k)}\nonumber\\
&&\qquad=(4\pi t)^{-\frac{d}{2}}\int_{(\R^d)^2}e^{-t\,V(x)}\,dx\label{Eqn:CrucialIneq}
\end{eqnarray*}
using
\[
\int_{(\R^d)^{n-1}}\kern -25pt dx_0\,dx_1\,dx_2\ldots\,dx_{k-1}\,dx_{k+1}\ldots\,dx_n\,\prod_{j=0}^n \!G\!\left(\frac tn,x_j-x_{j+1}\!\right)=G(t,x_k-x_k)=(4\pi t)^{-\frac{d}{2}}.
\]
\finprf

\noindent{\sl Proof of Theorem \ref{Thm:Main}.\/} The definition of the $\Gamma$ function gives, for any $\gamma >0$ and $\lambda >0$,
\[
\lambda^{-\gamma} = \frac{1}{\Gamma(\gamma)}\int_0^{+\infty} e^{-t\lambda}\,t^{\gamma-1}dt\, .
\]
The operator $-\Delta+V$ is essentially self-adjoint on $L^2(\R^d)$, and positive, since $V$ is nonnegative. This implies, by the functional calculus,
\[
\tr\left((-\Delta +V)^{-\gamma}\right) = \frac{1}{\Gamma(\gamma)}\int_0^{+\infty} \tr\left(e^{-t\,(-\Delta +V)}\right)\,t^{\gamma-1}dt\,.
\]
Using (\ref{Ineq:Symanzik}), since $V^{\frac{d}{2}-\gamma} \in L^1(\R^d)$, we get
\begin{eqnarray*}
\tr\left((-\Delta +V)^{-\gamma}\right) &\leq& \frac{1}{\Gamma(\gamma)}\int_0^{+\infty}\int_{\R^d}(4 \pi t)^{-\frac{d}{2}}e^{-tV(x)}\,t^{\gamma -1}dx \;dt \\
&\leq & \frac{\Gamma(\gamma - \frac{d}{2})}{(4\pi)^{\frac{d}{2}}\Gamma(\gamma)}\int_{\R^d}\left[V(x)\right]^{\frac{d}{2}-\gamma}dx\,.
\end{eqnarray*}
We define $\cal{C}(\gamma):=$ $\frac{\Gamma(\gamma - \frac{d}{2})}{(4\pi)^{\frac{d}{2}}\Gamma(\gamma)}$ and obtain the announced inequality.

The optimality of the constant is established by the following example. Consider the potential $V_\varepsilon\equiv 1$ in $(0,\varepsilon^{-1}\,\pi)^d=\Omega_\varepsilon\subset\R^d$, $V_\varepsilon\equiv +\infty$ in $\Omega_\varepsilon^c$. Such a potential can be approximated by smooth potentials $V_\varepsilon^n$ such that $V_\varepsilon^n\equiv 1$ in $\Omega_\varepsilon$ and $\lim_{n\to\infty}V_\varepsilon^n(x)=+\infty$ for any $x\in\Omega_\varepsilon^c$. The eigenvalues of $-\Delta +V_\varepsilon$ on $\R^d$ are the same as the ones of $-\Delta +V_\varepsilon$ on~$\Omega_\varepsilon$ with zero Dirichlet boundary conditions on $\partial \Omega_\varepsilon$:
\[1+\varepsilon^2\,\sum_{j=1}^d n_j^2\;,\quad n_1,\;n_2\,\ldots\,n_d\in\N^*\;,\]
so that
\[\tr\left((-\Delta +V_\varepsilon)^{-\gamma}\right)=\sum_{n_1,\;n_2\,\ldots\,n_d\in\N^*}\left(1+\varepsilon^2\,\sum_{j=1}^d n_j^2\right)^{-\gamma},\]
which behaves asymptotically as $\varepsilon$ tends to zero as 
\begin{eqnarray*}
\sum_{n_1,\;n_2\,\ldots\,n_d\in\N^*}\begin{array}{c} \\ \displaystyle\int\kern -5pt\int\ldots\int\\{\begin{array}{c}\scriptstyle n_j-1\leq x_j\leq n_j\\\scriptstyle j=1,2,\ldots d\end{array}}\end{array} \frac{dx}{\left(1+\varepsilon^2\,|x|^2\right)^\gamma}&=&
\frac 1{(2\,\varepsilon)^d} \int_{\R^d}\frac{dx}{\left(1+|x|^2\right)^\gamma}
\\
&=& \frac{|S^{d-1}|}{(2\,\varepsilon)^d}\int_0^\infty\frac{r^{d-1}}{(1+r^2)^\gamma}\;dr\;.
\end{eqnarray*}
This is precisely the right hand side of Inequality \eqn{main} as can be checked using $(\pi/\varepsilon)^d=\ix{V_\varepsilon^{d/2-\gamma}}$ and
\[
|S^{d-1}|=\frac{2\,\pi^{\frac{d}{2}}}{\Gamma(\frac{d}{2})}\quad\mbox{and}\quad \int_0^\infty{\frac{r^{d-1}}{{\left( 1 + r^2 \right) }^\gamma}}\;dr =\frac{\Gamma(\gamma - \frac{d}{2})\,\Gamma(\frac{d}{2})}{2\,\Gamma(\gamma)}\;.
\]
\finprf

{}From the above proof, it easy to see that the result of Theorem \ref{Thm:Main} can be generalized as follows. Let $f$ be a nonnegative function on $\R_+$ such that
\be{Eqn:AssGen}
\int_0^\infty f(t)\,\left(1+t^{-d/2}\right)\;\frac{dt}t<\infty
\ee
and define
\be{Eqn:FG}F(s):=\int_0^\infty e^{-t\,s}\,f(t)\;\frac {dt}t\quad\mbox{and}\quad G(s):=\int_0^\infty e^{-t\,s}\,\left(4\pi\,t\right)^{-d/2}\,f(t)\;\frac {dt}t\;.
\ee
Notice that if $d$ is even, $(-4\pi)^{(d/2)}\,d^{(d/2)}G/ds^{(d/2)}(s)=F(s)$. In the case of Theorem~\ref{Thm:Main}, $F(s)=s^{-\gamma}$ and $G(s)=\frac{\Gamma(\gamma - \frac{d}{2})}{(4\pi)^{\frac{d}{2}}\Gamma(\gamma)}\,s^{\frac{d}{2}-\gamma}$.
%-----------------------------------------------------------------------------
\begin{thm}\label{Thm:Main-Gen} Let $V$ be in $L^1_{\rm loc}(\R^d)$ and bounded from below. Assume moreover that $G(V)$ is in $L^1(\R^d)$. With $F$ and $G$ defined by \eqn{Eqn:FG}, if $f$ satisfies Asssumption \eqn{Eqn:AssGen}, then
\[
\sum_{i\in\N^*}F(\lambda_i(V))=\tr \left[ F\left(-\Delta +V\right)\right]\leq\int_{\R^d}G(V(x))\,dx\;.
\]\end{thm}
%-----------------------------------------------------------------------------
\proof The above inequality follows from the definition of $F$:
\[
\tr \left[ F\left(-\Delta +V\right)\right]=\int_0^{+\infty} \tr\left(e^{-t\,(-\Delta +V)}\right)\,f(t)\;\frac {dt}t\;,
\]
Inequality \eqn{Ineq:Symanzik} and the definition of $G$.\finprf

As an example, if we apply Theorem~\ref{Thm:Main-Gen} with $F(s)=e^{-s}$, $f(s)=\delta(s-1)$ and $G(s)=(4\pi)^{-d/2}\,e^{-s}$, we get
\be{Ineq:Exp}\sum_{i\in\N^*}e^{-\lambda_i(V)}\leq \frac 1{(4\pi)^{d/2}}\,\int_{\R^d}e^{-V(x)}\,dx\;.\ee
In the special case $V(x)=A^2\,|x|^2+B$, eigenvalues are explicitly given as
\[B+\sum_{j=1}^d\left(2\,n_j+1\right)\,A\;,\quad n_1,\;n_2\,\ldots\,n_d\in\N\;,\]
and we can compute
\be{Eqn:QuadPot}\tr\left(e^{-t\,(-\Delta +V)}\right)=\sum_{i\in\N^*}e^{-t\,\lambda_i(V)}=e^{-B\,t}\;\prod_{j=1}^d\left(\sum_{n_j\in\N^*}e^{-\left(2\,n_j+1\right)\,A\,t}\right)=\frac{e^{-B\,t}}{[2\,\sinh(At)]^d}\;.\ee
On the other hand, 
\[\frac 1{(4\pi)^{d/2}}\,\int_{\R^d}e^{-V(x)}\,dx=\frac {e^{-B}}{(2\,A)^d}\;.\]
Putting these estimates together in the case $t=1$ shows that the upper bound in \eqn{Ineq:Exp}, namely
\[\frac 1{(4\pi)^{d/2}}\,\left(\frac A{\sinh A}\right)^d=\frac{\sum_{i\in\N^*}e^{-\lambda_i(V)}}{\int_{\R^d}e^{-V(x)}\,dx}\le \frac 1{(4\pi)^{d/2}}\;,\]
is achieved in the limit $A\to 0_+$.

\medskip Identity \eqn {Eqn:QuadPot} is also useful in the case $F(s)=s^{-\gamma}$ considered in Theorem 1. Using $\tr\left((-\Delta +V)^{-\gamma}\right) = \frac{1}{\Gamma(\gamma)}\int_0^{+\infty} \tr\left(e^{-t\,(-\Delta +V)}\right)\,t^{\gamma-1}dt$, we obtain in the special case $V(x)=A^2\,|x|^2+B$ the identity
\[\tr\left((-\Delta +V)^{-\gamma}\right)=\frac{1}{\Gamma(\gamma)}\int_0^{+\infty}\kern-5pt \frac{e^{-B\,t}}{[2\,\sinh(At)]^d}\,t^{\gamma-1}\;dt=\frac{B^{-\gamma}}{\Gamma(\gamma)}\int_0^{+\infty}\kern-5pt \frac{e^{-t}}{[2\,\sinh(s\,t)]^d}\,t^{\gamma-1}\;dt\]
with $s:=B/A$. Under the additional restriction $\gamma >d$, we get
\[
\ix{V^{\frac d2-\gamma}}=B^{d-\gamma}\,A^{-d}\,\pi^{\frac d2}\,\frac{\Gamma(\gamma-d)}{\Gamma\left(\gamma-\frac d2\right)}\;.
\]
With $\CLTd=\frac{\Gamma(\gamma - \frac{d}{2})}{(4\pi)^{d/2}\Gamma(\gamma)}$, this shows that
\[
\frac{\tr\left((-\Delta +V)^{-\gamma}\right)}{{\cal C}(\gamma)\,\ix{V^{\frac d2-\gamma}}}=\frac{s^d}{\Gamma(\gamma-d)}\int_0^\infty\kern-5pt \frac{t^{\gamma-1}\,e^{-t}}{\left(\sinh (s\,t)\right)^d}\;dt=:q(s)\;.\]
It is easy to check that the function $s\mapsto q(s)$ bounded by $1$ and achieves $1$ in the limit $s\to 0_+$.

%%%%%%%%%%%%%%%%%%%%%%%%%%%%%%%%%%%%%%%%%%%%%%%%%%%%%%%%%%%%%%%%%%%%%%
%%%%%%%%%%%%%%%%%%%%%%%%%%%%%%%%%%%%%%%%%%%%%%%%%%%%%%%%%%%%%%%%%%%%%%
\section{Stability for the linear Schr\"odinger equation}\label{Sec:Stability}

In this section we come back to the physical motivation of Theorems~\ref{Thm:Main} and \ref{Thm:Main-Gen} with more details than in the introduction and state a list of examples corresponding to various functions $F$.

%%%%%%%%%%%%%%%%%%%%%%%%%%%%%%%%%%%%%%%%%%%%%%%%%%%%%%%%%%%%%%%%%%%%%%
\subsection{Notations and assumptions}
Let $E[\psi]:=\int_{\R^d}(|\nabla\psi|^2+V\,|\psi|^2)\, dx$ and assume that $V$ is a potential such that the operator $H_V:=-\Delta+V$ has an infinite nondecreasing sequence of eigenvalues $\seq{\lambda_i(V)}$:
\[
\lambda_i(V):=\inf_{\begin{array}{c}F\subset L^2(\R^d)\\ {\rm dim}(F)=i\end{array}} \sup_{\psi\in F} E[\psi]\;.
\]
Here the eigenvalues are counted with multiplicity, and to each $\lambda_i(V)$, $i\in\N^*$, we can associate an eigenfunction $\bar\psi_i$ such that $\bar\bpsi:=\seq{\bar\psi_i}$ is an orthonormal sequence:
\[
\scal{\bar\psi_i}{\bar\psi_j}=\delta_{ij}\quad\forall\; i,\, j\in\N^*\,.
\]
As in Section~\ref{Sec:Intro}, we also define $\bar\nu_i:=(\beta')^{-1}(-\lambda_i(V))$ for any $i\in\N^*$, $\bar\bnu:=\seq{\bar\nu_i}$. The {\sl free energy\/} of the {\sl mixed state\/} $(\bnu,\bpsi)=(\seq{\nu_i},\seq{\psi_i})\in\Sp$ is
\[
\F[\bnu,\bpsi]:= \sum_{i\in\N^*}\beta(\nu_i)+\sum_{i\in\N^*}\nu_i\,E[\psi_i]
\]
for some given function $\beta$. If the potential $V$ is such that $-\Delta +V$ has an unbounded sequence of eigenvalues, it is easy to see that $\F[\bnu,\bpsi]$ is defined only if $\lim_{i \rightarrow \infty} \nu_i = 0$. This allows us to re-order the sequence $(\bnu,\bpsi)$ in such a way that $(\nu_i)_{i \in \N^*}$ is a non increasing sequence converging to $0$, and we may restrict the domain of the free energy $\F$ to $\Spp$, 
where $S$ denotes the set of non increasing sequences in $\R_+$ converging 
to $0$, such that $\sum_{i\in\N^*}\beta(\nu_i)$ is absolutely convergent. Notice that whenever it is finite, $\sum_{i\in\N^*}\beta(\nu_i)$ is absolutely convergent by the assumption $\lim_{i \rightarrow \infty} \nu_i = 0$.

We shall say that {\sl Assumption\/} {\rm (H)} holds if $\beta$ is a strictly convex function with $\beta(0)=0$, which is $C^1$ on the interior of its support and if the potential $V$ is such that $-\Delta +V$ has an unbounded sequence of eigenvalues $\seq{\lambda_i(V)}$ for which
$$
\left|\sum_{i\in\N^*}\beta(\bar\nu_i)\right|<\infty\quad\mbox{and}\quad \left|\sum_{i\in\N^*}\bar\nu_i\,\lambda_i(V)\right|<\infty\;,
$$
where $\bar\nu_i:=(\beta')^{-1}(-\lambda_i(V))$ for any $i\in\N^*$. As seen in Section~\ref{Sec:Intro}, Assumption (H) is a consequence of the Lieb-Thirring type inequalities of Theorem~\ref{Thm:Main} if $\beta(\nu)=-(1-m)^{m-1}m^{-m}\,\nu^m$, $m\in (0,1)$ (see below Example 2 for more details). In the framework of Theorem~\ref{Thm:Main-Gen}, $F(\lambda)=-\beta(\nu)-\lambda\,\nu$ with $\nu=(\beta')^{-1}(-\lambda)$.

%%%%%%%%%%%%%%%%%%%%%%%%%%%%%%%%%%%%%%%%%%%%%%%%%%%%%%%%%%%%%%%%%%%%%%
\subsection{Minimizers of the free energy}

%---------------------------------------------------------------------
\begin{prop}\label{Lem:1} Assume that $\beta$ and $V$ are such that {\rm (H)} holds. Then there exists a minimizer $(\bar\bnu,\bar\bpsi)\in\Spp$ of $\F$ under the constraint
\[
\scal{\bar\psi_i}{\bar\psi_j}=\delta_{ij}\quad\forall\; i\,,\; j\in\N^*\,.
\]
Moreover,
\[\label{Eq:OptimalNu}
\bar\nu_i=(\beta')^{-1}(\lambda_i(V))
\]
and if $\bar\nu_i$ is positive for any $i\in\N^*$, the sequence $\bar\bpsi=\seq{\bar\psi_i}$ is unique up to any unitary transformation which leaves all eigenspaces of $-\Delta +V$ invariant. In particular, any $\bar\psi_i$ can be multiplied by an arbitrary constant phase factor, so that we may assume that $\bar\psi_i$ is real. \end{prop}
%---------------------------------------------------------------------
To prove this result we first prove some results about finite mixed states : 
given any $n\in\N^*$, we can define the projection $P_n$ of a mixed state $(\bnu,\bpsi)\in\Spp$ onto the $n$-finite mixed states by $P_n[\bnu,\bpsi]:=(\tilde\bnu,\bpsi)$ with $\tilde\nu_i=\nu_i$ for any $i=1,\,2,\ldots\,n$ and  $\tilde\nu_i=0$ for any $i\geq n+1$. Let $\F_n:=\F\circ P_n$:
$$
\F_n[\bnu,\bpsi]:= \sum_{i=1}^n\Big(\beta(\nu_i)+\nu_i\,E[\psi_i]\Big)\,.
$$
Notice indeed that $\beta(0)=0$, so that $\beta(\tilde\nu_i)=0$ for any $i\geq n+1$. We may decompose $\mathcal F_n$ into an entropy and an energy term as follows.
%---------------------------------------------------------------------
\begin{lemma}\label{Lem:2} Under Assumption {\rm (H)}, for any $(\bnu,\bpsi)\in\Spp$ such that $\bpsi=\seq{\psi_i}$ is an orthonormal sequence, 
\[
\F_n[\bnu,\bpsi]-\F_n[\bar\bnu,\bar\bpsi]=\sum_{i=1}^n\Big(\beta(\nu_i)-\beta(\bar\nu_i)-\beta'(\bar\nu_i)(\nu_i-\bar\nu_i)\Big)+\sum_{i=1}^n\nu_i\,\Big(E[\psi_i]-E[\bar\psi_i]\Big)\,.
\]
\end{lemma}
%---------------------------------------------------------------------
\proof An elementary computation gives
\[
\beta'(\bar\nu_i)\,(\nu_i-\bar\nu_i)+\nu_i\,E[\bar\psi_i]=-\lambda_i(V)\,(\nu_i-\bar\nu_i)+\nu_i\,\lambda_i(V)=\bar\nu_i\,\lambda_i(V)=\bar\nu_i\,E[\bar\psi_i]\;.
\]
\finprf

We are now going to study independently the two terms of $\F_n[\bnu,\bpsi]-\F_n[\bar\bnu,\bar\bpsi]$ and start with the entropy term.
%---------------------------------------------------------------------
\begin{lemma}\label{Lem:CK} Assume that $\inf_{s>0}\beta''(s)\,s^{2-p}=:\alpha>0$ for some $p\in[1,2]$. Then for any sequence $\seq{\nu_i}\in\R_+^{\N^*}$, if $\sum_{i\in\N^*}\beta(\nu_i)$ and $\sum_{i\in\N^*}\nu_i\,\beta'(\bar\nu_i)$ are absolutely convergent, then $\seq{\nu_i-\bar\nu_i}\in\ell^p$ and
\[
\sum_{i\in\N^*}\Big(\beta(\nu_i)-\beta(\bar\nu_i)-\beta'(\bar\nu_i)(\nu_i-\bar\nu_i)\Big)\geq 2^{-2/p}\,\alpha\,\|\bnu-\bar\bnu\|_{\ell^p}^2\cdot\min\left\{\|\bnu\|_{\ell^p}^{p-2},\|\bar\bnu\|_{\ell^p}^{p-2}\right\}\;.
\]
\end{lemma}
%---------------------------------------------------------------------
See \cite{Caceres-Carrillo-Dolbeault03} for a continuous version of this inequality. We may also refer to \cite{MR634765} in the case of $\beta(\nu)=\nu\log\nu-\nu$ and von Neumann algebras, and to \cite{MR1801751} for a review of the so-called Csisz\'ar-Kullback inequalities. For the completeness of the paper, we give a short proof of this result.\smallskip

\proof For any $i\in\N^*$, let $\zeta_i\in[\min(\nu_i,\bar\nu_i),\max(\nu_i,\bar\nu_i)]$ be an intermediate nonnegative point such that 
\[
\sum_{i\in\N^*}\Big(\beta(\nu_i)-\beta(\bar\nu_i)-\beta'(\bar\nu_i)(\nu_i-\bar\nu_i)\Big)=\frac 12\sum_{i\in\N^*}\beta'(\zeta_i)(\nu_i-\bar\nu_i)^2\geq \frac \alpha{2}\sum_{i\in\N^*}\zeta_i^{p-2}(\nu_i-\bar\nu_i)^2\;.\]
Let ${\mathcal I}\subset\N^*$. Using
\[
\left(\sum_{i\in{\mathcal I}}|\nu_i-\bar\nu_i|^p\,\zeta_i^{p(p-2)/2}\cdot \zeta_i^{p(2-p)/2}\right)^{2/p}\leq \sum_{i\in{\mathcal I}}\zeta_i^{p-2}(\nu_i-\bar\nu_i)^2\cdot\left(\sum_{i\in{\mathcal I}}\zeta_i^p\right)^{(2-p)/p},
\]
we get 
\[
\sum_{i\in{\mathcal I}}\zeta_i^{p-2}(\nu_i-\bar\nu_i)^2\geq \left(\sum_{i\in{\mathcal I}}|\nu_i-\bar\nu_i|^p\right)^{2/p}\cdot\left(\sum_{i\in{\mathcal I}}\zeta_i^p\right)^{1-2/p}.
\]
On ${\mathcal I}=\{i\in\N^*\,:\,\nu_i>\bar\nu_i\}$ (respectively ${\mathcal I}=\{i\in\N^*\,:\,\nu_i<\bar\nu_i\}$), we estimate $\sum_{i\in{\mathcal I}}\zeta_i^p$ from above by $\sum_{i\in{\mathcal I}}\nu_i^p$ (resp. by $\sum_{i\in{\mathcal I}}\bar\nu_i^p$). Using the inequality $(a+b)^r\le 2^{r-1}(a^r+b^r)$ for any $a$, $b\geq 0$ and $2/p=r\geq 1$, we completes the proof. \finprf

Next, we turn our attention to the energy term and recall a result given, for instance, in \cite{MR2001i:00001}: 
%---------------------------------------------------------------------
\begin{prop}\label{minmaxloss}
Let $V$ be a potential such that the sequence of eigenvalues $\seq{\lambda_i(V)}$ of $H_V$ is unbounded, and choose any $n$ functions $\psi_1,\dots,\psi_n$ that are orthonormal in $L^2(\R^d)$. Then 
$$
\sum_{i=1}^n E[\psi_i] \geq \sum_{i=1}^n \lambda_i(V)\;.
$$
\end{prop}
%---------------------------------------------------------------------
We extend this property to orthogonal families:
%---------------------------------------------------------------------
\begin{lemma}\label{lempatu1} Assume that $V$ is a potential as above. 
For any orthogonal family $(\phi_i)_{1\leq i \leq n}$ in $L^2(\R^d)$, with $
\|\phi_i\|^2 = \nu_i$ and $\nu_1 \geq \dots \geq \nu_n$, we get 
$$
\sum_{i=1}^n E[\phi_i] \geq \sum_{i=1}^n \nu_i\,\lambda_i(V)\;.
$$
\end{lemma}
%---------------------------------------------------------------------
\proof We prove this result by induction on $n$. The case $n=1$ is trivial. Suppose that the result holds for any orthogonal system of $n$ functions, and take $(\phi_i)_{1\leq i \leq n+1}$ with nonincreasing squared norms (or occupation numbers) $\nu_1 \geq \dots \geq \nu_{n+1}\geq 0$. If $\nu_{n+1}=0$, then the induction assumption directly gives the result. Assume next that $\nu_{n+1}>0$. By Proposition~\ref{minmaxloss}, we have
$$
\sum_{i=1}^{n+1} E\left[\frac{\phi_i}{\|\phi_i\|}\right] \geq \sum_{i=1}^{n+1} \lambda_i(V)\;.
$$
Multiplying by $\nu_{n+1}$, we obtain
$$
\sum_{i=1}^{n+1}\frac{\nu_{n+1}}{\nu_i}\,E[\phi_i] \geq \sum_{i=1}^{n+1} \nu_{n+1}\,\lambda_i(V)\;,
$$
hence
\begin{equation}\label{crochet}
\sum_{i=1}^{n+1}\,E[\phi_i] \geq \sum_{i=1}^{n} \left[ \frac{\nu_i-\nu_{n+1}}{\nu_i}\,E[\phi_i] - (\nu_i-\nu_{n+1})\,\lambda_i(V)\right] + \sum_{i=1}^{n+1}\nu_i\,\lambda_i(V)\,.
\end{equation}
Since $\nu_i \geq \nu_{n+1}$, we can define the family $(\mu_i\,\phi_i)_{1 \leq i\leq n}$ with $\mu_i:=(\frac{\nu_i-\nu_{n+1}}{\nu_i})^{1/2}$, which is orthogonal. By the induction hypothesis  we get
$$
\sum_{i=1}^{n}\frac{\nu_i-\nu_{n+1}}{\nu_i}\,E[\phi_i] = \sum_{i=1}^{n} E \left(\mu_i\,\phi_i\right) \geq \sum_{i=1}^{n}\left\|\mu_i\,\phi_i\right\|^2 \lambda_i(V) = \sum_{i=1}^{n}(\nu_i- \nu_{n+1})\,\lambda_i(V)\;.
$$
In Inequality (\ref{crochet}), the first sum of the right hand side is then nonnegative. For the system of the $n+1$ orthogonal functions, we obtain
$$ \sum_{i=1}^{n+1} E[\phi_i] \geq \sum_{i=1}^{n+1}\nu_i\,\lambda_i(V)\,,$$
which completes the proof of Lemma \ref{lempatu1}.\finprf

\noindent{\sl Proof of Proposition \ref{Lem:1}.\/}  By Lemma \ref{lempatu1} we get
$$
\F_n[\bnu,\bpsi] \geq \sum_{i=1}^n\Big(\beta(\nu_i)+\nu_{i}\,\lambda_{i}(V)\Big),
$$ hence a  minimization of $\F_n$ under the constraint $\scal{\psi_i}{\psi_j}=\delta_{ij}$ directly gives, for any $[\bnu,\bpsi] \in \Spp$,
$$ \F_n[\bnu,\bpsi] \geq \F_n[\bar\bnu,\bar\bpsi]\,.$$

Assumption (H) gives the absolute convergence of the series in the definition of $\F(\bar\bnu,\bar\bpsi)$. Suppose now that there exists $(\tilde{\bnu},\tilde{\bpsi})\in\Spp$ such that $\F(\tilde{\bnu},\tilde{\bpsi}) < \F(\bar\bnu,\bar\bpsi)$. This implies the existence of a $N \in \N^*$ such that 
$$ \sum_{i=1}^N\Big(\beta(\tilde{\nu_i})+\tilde{\nu_{i}}\,E[\tilde{\psi_{i}}]\Big) < \sum_{i=1}^N\Big(\beta(\bar\nu_i)+\bar\nu_{i}\,E[\bar\psi_{i}]\Big)\,,$$
which contradicts the result on finite mixed states.\finprf

%%%%%%%%%%%%%%%%%%%%%%%%%%%%%%%%%%%%%%%%%%%%%%%%%%%%%%%%%%%%%%%%%%%%%%
\subsection{Stability}

As a consequence of the conservation of the energy $E[\cdot]$ under the evolution according to the Schr\"odinger operator $i\,\partial_t-H_V$, we obtain the conservation of the free energy. Notice here that all above computations have been done with functions taking real values and need to be adapted to the case of complex valued functions as soon as we consider solutions to the time-dependent Schr\"odinger equation.
%---------------------------------------------------------------------
\begin{prop}\label{Cor:Cor1} Assume {\rm (H)} and consider an initial mixed state $(\bnu,\bpsi^0)\in\Sp$. If $(\bnu,\bpsi(t))$ is the mixed state where each of the components evolves according to the linear Schr\"odinger equation
\[\label{Eq:Schroedinger}
i\,\partial_t\psi_j=-\Delta\psi_j+V\psi_j\;,\quad x\in\R^d\,,\; t>0
\]
with initial data $\psi_j^0$ for any $j\in\N^*$, then 
\[
\F[\bnu,\bpsi(t)]=\F[\bnu,\bpsi^0]\quad\forall\;t>0\;.
\]\end{prop}
%---------------------------------------------------------------------
To state a dynamical stability result, we have to impose a decay property of the sequence of occupation numbers as in Lemma~\ref{lempatu1}. From Lemmas~\ref{Lem:CK} and \ref{lempatu1}, and Proposition~\ref{Cor:Cor1}, we deduce the
%---------------------------------------------------------------------
\begin{corollary}\label{Cor:Cor2} Consider an initial mixed state $(\bnu,\bpsi^0)\in\Spp $ with a nonincreasing sequence of occupation numbers $\bnu$. Under the assumption of Lemma \ref{Lem:CK},  if {\rm (H)} is satisfied, then for any $t>0$,
\[
2^{-2/p}\,\alpha\,\|\bnu-\bar\bnu\|_{\ell^p}^2\cdot\min\left\{\|\bnu\|_{\ell^p}^{p-2},\|\bar\bnu\|_{\ell^p}^{p-2}\right\}+\sum_{i\in\N^*}\nu_i\,\Big(E[\psi_i(t)]-\lambda_i(V)\Big)\leq \F[\bnu,\bpsi^0]\;,
\]
where both terms of the left hand side are nonnegative.\end{corollary}
%---------------------------------------------------------------------

%%%%%%%%%%%%%%%%%%%%%%%%%%%%%%%%%%%%%%%%%%%%%%%%%%%%%%%%%%%%%%%%%%%%%%
\subsection{Examples}

We conclude these comments on stability results by a list of examples of various functions $\beta$ and by the corresponding Lieb-Thirring type inequalities given by Theorem~\ref{Thm:Main-Gen} with $-F(s)=(\beta\circ(\beta')^{-1})(-s)+s\,(\beta')^{-1}(-s)$. We refer to \cite{MR1847430,DOMS} for a similar discussion in a non quantum mechanical context.

\par\medskip\noindent\underline{{\sl Example 1.\/}} Let $m>1$ and consider $\beta(\nu):=(m-1)^{m-1}m^{-m}\,\nu^m$. With $\beta'(\nu)=(m-1)^{m-1}m^{1-m}\,\nu^{m-1}=-\lambda$ and $m=\frac\gamma{\gamma-1}$, we get: $-(\beta(\nu)+\lambda\,\nu)=F(\lambda)=(-\lambda)^\gamma$, which corresponds to the setting of the standard Lieb-Thirring inequality~\eqn{Ineq:LT}. The case $\gamma\in (0,1)$ is formally covered by $\beta(\nu):=-(1-m)^{m-1}|m|^{-m}\,\nu^m$ with $m\in (-\infty,0)$, $m=\frac\gamma{\gamma-1}$ again and $F(s)=(-s)^\gamma$, but in this case, $\beta$ is not convex and the free energy $\F$ cannot be defined as above. 

\par\medskip\noindent\underline{{\sl Example 2.\/}} As seen above, for $m<1$ and $\beta(\nu):=-(1-m)^{m-1}m^{-m}\,\nu^m$, with $\beta'(\nu)=-(1-m)^{m-1}m^{1-m}\,\nu^{m-1}=-\lambda$ and $m=\frac\gamma{\gamma+1}$, we get: $F(\lambda)=\lambda^{-\gamma}$, which corresponds to the setting of Theorem~\ref{Thm:Main}.

\par\medskip\noindent\underline{{\sl Example 3.\/}} If $\beta(\nu):=\nu\log\nu-\nu$, then $\beta'(\nu)=\log\nu=-\lambda$. According to Theorem~\ref{Thm:Main-Gen}, the corresponding inequality is
\[
\sum_{i\in\N^*}e^{-\lambda_i(V)}\leq\frac 1{(4\pi)^{d/2}}\ix{e^{-V(x)}}\;.
\]
This case can formally be seen as the limit case $m\to 1$ in Examples 1 and 2. Here $F(s)=e^{-s}$, $G(s)=(4\pi)^{-d/2}\,e^{-s}$.

\par\medskip\noindent\underline{{\sl Example 4.\/}} If $\beta(\nu):=\nu\log\nu+(1-\nu)\log(1-\nu)$, then $\beta'(\nu)=\log\big(\frac\nu{1-\nu}\big)=-\lambda$ and $F(s)=\log(1+e^{-s})$. According to Theorem~\ref{Thm:Main-Gen}, the corresponding inequality is
\[
\sum_{i\in\N^*}\log\left(1+e^{-\lambda_i(V)}\right) \leq \ix{G(V(x))}
\]
where $G$ is given in terms of $F$ by \eqn{Eqn:FG}. 

\medskip In all the above examples we have to assume that $\lim_{i\to\infty}\lambda_i(V)=+\infty$, except in Example 1 where $\lambda_i(V)\leq 0$, $\lim_{i\to\infty}\lambda_i(V)=0$ and we adopt the convention that $\lambda_i(V)=0$ for any $i>N$ if there are only $N$ negative eigenvalues.

%%%%%%%%%%%%%%%%%%%%%%%%%%%%%%%%%%%%%%%%%%%%%%%%%%%%%%%%%%%%%%%%%%%%%%
%%%%%%%%%%%%%%%%%%%%%%%%%%%%%%%%%%%%%%%%%%%%%%%%%%%%%%%%%%%%%%%%%%%%%%
\section{Lieb-Thirring Gagliardo-Nirenberg inequalities}\label{Sec:LT-GN}

In this section, we will focus on consequences of Theorems~\ref{Thm:Main} and \ref{Thm:Main-Gen}, when one takes only partial sums, and especially when only the first eigenvalue is considered. 

%%%%%%%%%%%%%%%%%%%%%%%%%%%%%%%%%%%%%%%%%%%%%%%%%%%%%%%%%%%%%%%%%%%%%%
\subsection{Optimal constant in the Lieb-Thirring conjecture}\label{Sec:Lieb-ThirringConjecture}

We begin with a remark on the connection of the best constant in the Lieb-Thirring conjecture and its extension for $d>1$:
\[
\CLTc:=\inf_{\begin{array}{c} V\in{\cal D}(\R^d)\\ V\leq 0\end{array}} \frac{|\lambda_1(V)|^\gamma}{\int_{\R^d}|V|^{\gamma+\frac d2}\;dx}
\]
with the best constant in some special Gagliardo-Nirenberg inequalities. Such a result has already been established in \cite{Benguria-Loss,MR2003g:35039} (also see \cite{MR84h:81097b,MR84h:81097a,MR84m:81006} for earlier references). We give it here for completeness and in order to insist on some interesting scaling properties.

Define the function set for the potential $V$ by 
\[
X_\gamma:=\left\{V\in L^{\gamma+\frac d2}(\R^d)\;:\; V\leq0\,,\;V\not\equiv 0\;a.e.\right\}
\]
and note that by density of ${\cal D}(\R^d)$ in $L^{\gamma+\frac d2}(\R^d)$, it holds that
\[
\CLTc\qquad =\sup_{\begin{array}{c} V\in X_\gamma\\
V\geq0,\;V\not\equiv 0\;a.e.\end{array}} \frac{|\lambda_1(V)|^\gamma}{\ix{V^{\gamma+\frac d2}}}\;.
\]
By density of ${\cal D}(\R^d)$ in $H^1(\R^d)$, we have
\[
\lambda_1(V)\qquad =\inf_{\begin{array}{c}u\in H^1(\R^d)\\u\not\equiv 0\;a.e.\end{array}}\frac{\ix{|\nabla u|^2}+\ix{V\,|u|^2}}{\ix{|u|^2}}\;.
\]
Let 
\[
q:=\frac {2\gamma+d}{2\gamma+d-2}
\]
and consider the optimal constant $\CGN$ of the Gagliardo-Nirenberg inequality corresponding to the embedding of $H^1(\R^d)$ into $L^{2q}(\R^d)$:
\be{Ineq:Gagliardo-Nirenberg1}
\CGN\qquad =\inf_{\begin{array}{c}u\in H^1(\R^d)\\ u\not\equiv 0\;a.e.\end{array}} \frac{\nrm{\nabla u}2^{\frac d{2\gamma+d}}\nrm{u}2^{\frac {2\gamma}{2\gamma+d}}}{\nrm{u}{2q}}\;.
\ee
Notice that for $\gamma>\max(0,1-d/2)$,
\[
q>1\quad\mbox{and}\quad 2\,q<\frac{2d}{d-2}\;.
\]
%---------------------------------------------------------------------
\begin{thm}\label{Thm:BestConstant} Let $d\in\N^*$. For any $\gamma>\max(0,1-\frac d2)$, 
\[\label{Eqn:BestConstant}
\CLTc=\kappa_1(\gamma)\,\Big[\CGN\Big]^{-\kappa_2(\gamma)}\;,
\]
where
\[
\kappa_1(\gamma)=\frac{2\gamma}d\left(\frac d{2\gamma+d}\right)^{1+\frac d{2\gamma}}\quad\mbox{and}\quad\kappa_2(\gamma)=2+\frac d\gamma\;.
\]
Moreover, the constant $\CLTc$ is optimal and achieved by a unique pair of functions $(u,V)$, up to multiplications by a constant, scalings and translations.
\end{thm}
%---------------------------------------------------------------------
The scaling invariance can be made clear by redefining
\[
\Big[\CLTc\Big]^{\frac1\gamma}\qquad =\sup_{\begin{array}{c} V\in X_\gamma\\
V\geq0,\;V\not\equiv 0\;a.e.\end{array}}\,\sup_{\begin{array}{c}u\in H^1(\R^d)\\ u\not\equiv 0\;a.e.\end{array}}R(u,V)
\]
where 
\[
R(u,V)=-\frac{\ix{V\,|u|^2}+\ix{|\nabla u|^2}}{\ix{|u|^2}\quad\nrm{V}{\gamma+\frac d2}^{1+\frac d{2\gamma}}}\;.
\]
Note indeed that $\lambda_1(V)\leq 0$, and $R(u,V)$ is invariant under the transformation
\[
(u,V)\mapsto \Big(u_\lambda=u(\lambda\,\cdot), \;V_\lambda=\lambda^2V(\lambda\,\cdot)\Big)\;,
\]
{\sl i.e.\/}, $R(u_\lambda, V_\lambda)=R(u,V)$ for any $\lambda>0$.

\medskip\noindent{\sl Proof of Theorem \ref{Thm:BestConstant}.\/} By H\"older's inequality, 
\[
\ix{|V|\,|u|^2}\leq A\, \nrm{u}{2q}^2\quad\mbox{with}\quad A:=\nrm{V}{\gamma+\frac d2}\;.
\]
Let $\tau:=\nrm{u}{2q}/\nrm{u}2$. The Gagliardo-Nirenberg inequality \eqn{Ineq:Gagliardo-Nirenberg1}, namely
\[
\CGN\,\nrm{u}{2q}\leq\nrm{\nabla u}2^\theta\;\nrm{u}2^{1-\theta}
\]
with $\theta=\frac d{2\gamma+d}$ can be rewritten as 
\[
\frac{\nrm{\nabla u}2}{\nrm{u}2}\geq \Big[\CGN\,\tau\Big]^{\frac 1\theta}\;.
\]
Putting these estimates together, we obtain 
\[
R(u,V)\leq \frac{A\,\tau^2-[\CGN]^{\frac 2\theta}\,\tau^{\frac 2\theta}}{A^{1+\frac d{2\gamma}}}\;.
\]
An optimization on $\tau$ shows the bound of Theorem \ref{Thm:BestConstant}, which is independent of $A$, and gives the expressions of $\kappa_1(\gamma)$ and $\kappa_2(\gamma)$.

\smallskip The estimate is achieved since all above inequalities can be saturated by considering 
\be{Eqn:Euler-Lagrange1}
|V|^{\gamma+\frac d2-2}\,V=|u|^2\quad\Longleftrightarrow\quad V=V_u=-|u|^{\frac 4{2\gamma+d-2}}=|u|^{2(q-1)}\;,
\ee
where $u$ is a solution of 
\[
\Delta u+|u|^{2(q-1)}u-u=0\quad\mbox{in}\quad\R^d\;.
\]
Up to a scaling, these two equations are the Euler-Lagrange equations corresponding to the maximization in $V$ and $u$ respectively. Because of the second equation, the relation with the Gagliardo-Nirenberg inequality is straightforward. In other words, 
\[
R(u,V)\leq R(u,V_u)=\frac{\ix{|u|^{2q}}-\ix{|\nabla u|^2}}{\ix{|u|^2}\left(\ix{|u|^{2q}}\right)^{\frac 1\gamma}}=\frac{\ix{|u_\lambda|^{2q}}-\ix{|\nabla u_\lambda|^2}}{\left(\ix{|u_\lambda|^{2q}}\right)^{\frac 1\gamma}}
\]
where $u_\lambda=\lambda^{\frac 1{q-1}}u(\lambda\,\cdot)$ and $\lambda^{\left(\frac d2-\frac 1{q-1}\right)}=\nrm u2$, so that $\nrm{u_\lambda}2=1$, 
\[
R(u,V)\leq\quad\frac{\tau^{2q}-[\CGN]^{\frac 2\theta}\,\tau^{\frac 2\theta}}{\tau^{\frac{2q}\gamma}}\;,
\]
and the result holds by optimizing in $\tau=\nrm{u_\lambda}{2q}$.\finprf

%---------------------------------------------------------------------
\noindent{\bf Remark. }{\sl The optimal function in the Gagliardo-Nirenberg inequality \eqn{Ineq:Gagliardo-Nirenberg1} is given as the nonnegative solution of \eqn{Eqn:Euler-Lagrange1} in $H^1(\R^d)$. This solution is radial, positive, decreasing, and unique up to translations, multiplication by constants and scalings. See for instance \cite{MR1803216} for uniqueness results of radial solutions, and references therein for earlier related results. Optimal function are not explicitly known in general but are easy to compute numerically as well as the optimal constants, see for instance \cite{MR90m:35051, MR2003g:35039}.}
%---------------------------------------------------------------------

%%%%%%%%%%%%%%%%%%%%%%%%%%%%%%%%%%%%%%%%%%%%%%%%%%%%%%%%%%%%%%%%%%%%%%
\subsection{Theorem \ref{Thm:Main} and Gagliardo-Nirenberg inequalities}\label{Sec:DualCase}

In this section, we adapt the remarks of Section \ref{Sec:Lieb-ThirringConjecture} to the case $V\geq 0$ of Theorem \ref{Thm:Main}. The interpolation of $\nrm u{2q}$, with $1<q<d/(d-2)$, $d\geq 3$, between $\nrm{\nabla u}2$ and $\nrm u2$ of the previous section is a standard case of Gagliardo-Nirenberg inequalities, but there is also another interesting case in Gagliardo-Nirenberg inequalities, which is somewhat less standard. It corresponds to the interpolation of $\nrm u2$ between $\nrm{\nabla u}2$ and $\ix{|u|^{2q}}$ for some $q\in (0,1)$. See \cite{DelPino-Dolbeault01a} for a similar setting, where both cases have been taken into account. What we establish in this section is that these less standard inequalities are related to the estimate of $[\lambda_1(V)]^{-\gamma}$ in terms of $\ix{V^{d/2-\gamma}}$.

Consider now a nonnegative smooth potential $V\in{\cal C}^\infty(\R^d)$ such that
\[
\lim_{|x|\to+\infty}V(x)=+\infty
\]
and denote by $\lambda_1(V)$, $\lambda_2(V)$,\,\ldots the positive eigenvalues of $-\Delta+V$. By density we may extend this set of potentials to the set
\[
Y_\gamma:=\left\{V^{\frac d2-\gamma}\in L^1(\R^d)\;:\; V\geq0\,,\;V\not\equiv +\infty\;a.e.\right\}\;.
\]
Let
\[
q:=\frac{2\gamma-d}{2(\gamma+1)-d}\in (0,1)
\]
and define an optimal constant of a second type Gagliardo-Nirenberg inequality by
\be{Ineq:Gagliardo-Nirenberg2}
\CGNd\qquad =\inf_{\begin{array}{c}u\in H^1(\R^d),\, u\not\equiv 0\;a.e.\\ \int_{\R^d}|u|^{2q}\, dx<\infty\end{array}} \frac{\nrm{\nabla u}2^{\frac d{2\gamma}}\;\left(\ix{|u|^{2q}}\right)^{\frac 1{2q}(1-\frac d{2\gamma})}}{\nrm u2}\;.
\ee
%---------------------------------------------------------------------
\begin{thm}\label{Thm:DualBestConstant} Let $d\in\N^*$. For any $\gamma>d/2$, there exists a positive constant $\CLTcd$
such that, for any $V\in Y_\gamma$, 
\[\label{Ineq:Dual}
\left[\lambda_1(V)\right]^{-\gamma}\leq\CLTcd\,\ix{V^{\frac d2-\gamma}}\;.
\]
As in Theorem \ref{Thm:BestConstant}, the optimal value of $\CLTcd$ is such that 
\[\label{Eqn:DualBestConstant}
\CLTcd=\kappa_1(\gamma)\,\Big[\CGNd\Big]^{-\kappa_2(\gamma)}\;,
\]
where $\kappa_1(\gamma)=\frac{(2q)^{\gamma-\frac d2}(d(1-q))^{\frac d2}}{(d(1-q)+2q)^\gamma}\quad\mbox{and}\quad\kappa_2(\gamma)=2\gamma$. Moreover, the constant $\CLTcd$ is achieved by a unique pair of functions $(u,V)$, up to multiplications by a constant, scalings and translations.
\end{thm}
%---------------------------------------------------------------------
Notice that $q<1$, and $2q>1$ if and only if $\gamma>1+d/2$. The best constant in the above inequality is
\[
\CLTc\qquad :=\sup_{\begin{array}{c} V\in Y_\gamma\\
V\geq0,\;V\not\equiv 0\;a.e.\end{array}} \frac{[\lambda_1(V)]^{-\gamma}}{\ix{V^{\frac d2-\gamma}}}\;.
\]
The scaling invariance can be made clear by writing
\[
\Big[\CLTcd\Big]^{\frac1\gamma}\qquad =\sup_{\begin{array}{c} V\in X_\gamma\\
V\geq0,\;V\not\equiv 0\;a.e.\end{array}}\,\sup_{\begin{array}{c}u\in H^1(\R^d)\\ u\not\equiv 0\;a.e.\end{array}}R(u,V)
\]
where 
\[
R(u,V):=\frac{\ix{|u|^2}\quad\left(\ix{V^{\frac d2-\gamma}}\right)^{\frac 1\gamma}}{\ix{|\nabla u|^2}+\ix{V\,|u|^2}}
\]
is invariant under the transformation
\[
(u,V)\mapsto \Big(u_\lambda=u(\lambda\,\cdot), \;V_\lambda=\lambda^2V(\lambda\,\cdot)\Big)\;,
\]
{\sl i.e.\/}, $R(u_\lambda, V_\lambda)=R(u,V)$ for any $\lambda>0$.

\medskip\noindent{\sl Proof of Theorem \ref{Thm:DualBestConstant}.\/} By H\"older's inequality, 
\[
\ix{u^{2q}}=\ix{u^{2q}\,V^q\cdot V^{-q}}\leq\left(\ix{V\,|u|^2}\right)^q\left(\ix{V^{-\frac q{1-q}}}\right)^{1-q}\;.
\]
With $A:=\left(\ix{V^{-q/(1-q)}}\right)^{-(1-q)/q}=\left(\ix{V^{d/2-\gamma}}\right)^{2/(2\gamma-d)}$, this means that 
\[
\ix{V\,|u|^2}\geq A\,\left(\ix{|u|^{2q}}\right)^{\frac 1q}\;.
\]
We may therefore estimate $R(u,V)$ as follows:
\[
R(u,V)\leq\frac{\nrm u2^2\quad A^{1-\frac d{2\gamma}}}{\nrm{\nabla u}2^2 +A\,\left(\ix{|u|^{2q}}\right)^{\frac 1q}}\;.
\]
An optimization under the scaling $\lambda\mapsto\lambda^{-d/2}u(\cdot/\lambda)$, which leaves the $L^2(\R^d)$-norm invariant, shows that
\[
\nrm{\nabla u}2^2 +A\,\nrm u2^2\geq \nrm{\nabla u}{2q}^{\frac{2d(1-q)}{d(1-q)+2q}}\left(\ix{|u|^{2q}}\right)^{\frac 2{d(1-q)+2q}}\kern -2pt A^{\frac{2q}{d(1-q)+2q}}\;(\kappa_1(\gamma))^{-\frac 1\gamma}
\]
using $\frac{2q}{d(1-q)+2q}=1-\frac d{2\gamma}$. Using the Gagliardo-Nirenberg inequality \eqn{Ineq:Gagliardo-Nirenberg2}, we get
\[
\nrm{\nabla u}2^2 +A\,\nrm u2^2\geq \left|\CGNd\right|^2\;\nrm u2^2\;A^{1-\frac d{2\gamma}}\;(\kappa_1(\gamma))^{-\frac 1\gamma}
\]
which proves that $\CLTcd\leq\kappa_1(\gamma)\,[\CGNd]^{-\kappa_2(\gamma)}$. It is moreover easy to check that the equality holds in H\"older's inequality if $V^{\frac d2-\gamma-1}$ is proportional to $|u|^2$. By taking a minimizer of \eqn{Ineq:Gagliardo-Nirenberg2}, this completes the proof of Theorem~\ref{Thm:DualBestConstant}. \finprf

%---------------------------------------------------------------------
\noindent{\bf Remark. }{\sl Notice that optimal functions are not explicitly known, unless $d=1$. Solutions to the Euler-Lagrange equations have compact support and minimal ones are radially symmetric and unique up to translations, see \cite{MR1387457}. Also see \cite{DelPino-Dolbeault01a} for more details.}
%---------------------------------------------------------------------

%%%%%%%%%%%%%%%%%%%%%%%%%%%%%%%%%%%%%%%%%%%%%%%%%%%%%%%%%%%%%%%%%%%%%%
\subsection{General case}\label{Sec:General case}

We may try to generalize the approach used for power laws to general nonlinearities like the ones of Theorem~\ref{Thm:Main-Gen}. However, this is not as simple as when evident scaling properties are present. We may indeed write
\[
\CF=\sup_V\frac{F(\lambda_1(V))}{\ix{G(V(x))}}\leq 1\;,
\]
where the above supremum is taken on an appropriate space. Assuming that $F$ is nonincreasing, we may characterize $\CF$ as 
\[
\CF=\sup_{\begin{array}{c}V,\,\phi\\ \ix{|\phi|^2}=1\end{array}}\frac{\displaystyle F\left(\ix{\Big(|\nabla\phi|^2+V\,|\phi|^2\Big)}\right)}{\ix{G(V(x))}}\;,
\]
so that the optimal value is at least formally given by 
\[
\CF=\sup_{\begin{array}{c}\phi\in H^1(\R^d)\\ \ix{|\phi|^2}=1\end{array}}\frac{\displaystyle F\left(\ix{\Big(|\nabla\phi|^2+|\phi|^2\,(G')^{-1}(\kappa\,|\phi|^2)\Big)}\right)}{\ix{\left(G\circ(G')^{-1}\right)(\kappa\,|\phi|^2)}}
\]
where $\kappa$ is given in terms of $\phi$ by
\[
\kappa=\left(\CF\right)^{-1}\,{\displaystyle F'\left(\ix{\Big(|\nabla\phi|^2+|\phi|^2\,(G')^{-1}(\kappa\,|\phi|^2)\Big)}\right)}\;.
\]
This indeed results of the optimization with respect to $V$, which amounts to
\[
\kappa\,|\phi|^2-G'(V)=0\;.
\]

This strategy is however easy to implement in one more case: $F(s)=e^{-s}$. In this~case, 
\[
\CF=\sup_{\begin{array}{c}V,\,\phi\\ \ix{|\phi|^2}=1\end{array}}\frac{\displaystyle e^{-\ix{(|\nabla\phi|^2+V|\phi|^2)}}}{(4\pi)^{-d/2}\ix{e^{-V}}}\;.
\]
The optimization with respect to $V$ gives
\[
V=-\log(|\phi|^2)
\]
up to an additive constant such that $\ix{e^{-V}}=\ix{|\phi|^2}=1$, which plays no role because its contribution to the numerator and to the denominator cancel. Summing up the inequality is therefore simply equivalent to the usual logarithmic Sobolev inequality: for any $\phi\in H^1(\R^d)$ such that $\ix{|\phi|^2}=1$,
\[
\ix{|\phi|^2\log(|\phi|^2)}+\log\left(\frac{(4\pi)^{d/2}}{\CF}\right) \leq \ix{|\nabla\phi|^2}\;.
\]
{}From standard results on logarithmic Sobolev inequalities, see for instance \cite{MR1038450}, it is known that optimal functions $\phi$ are gaussian, which allows to determine the value of~ $\CF$:
\[
\CF=\left(\frac 2e\right)^d\,.
\]
We will see later an alternative approach which allows to state the following interpolation inequality.
%---------------------------------------------------------------------
\begin{prop}\label{Prop:Interpolation} Under the assumptions of Theorem~\ref{Thm:Main-Gen}, if $F$ and $G$ are related by \eqn{Eqn:FG}, if $F'$ and $G'$ are invertible and if we define
\[
\beta(s):=-\int_0^s (F')^{-1}(-t)\;dt\quad\mbox{and}\quad H(s):=\int_s^0(G')^{-1}(-t)\;dt\;,
\]
then for any $\phi\in H^1(\R^d)$, the following interpolation inequality holds:
\[
\ix{|\nabla\phi|^2}+\beta\left(\ix{|\phi|^2}\right)\geq\ix{H(|\phi|^2)}\;.
\]
\end{prop}
%---------------------------------------------------------------------
This result will appear as a simple consequence of Theorem~\ref{Thm:Interpolation}, where we take $\nu_1=\ix{|\phi|^2}$ and $\nu_i=0$ for any $i\geq 2$. We will see that the result holds not only in the framework of Theorem~\ref{Thm:Main-Gen} but also in the case where $\lim_{i\to\infty}\lambda_i(V)<\infty$ as it is the case for standard Lieb-Thirring inequalities.

%%%%%%%%%%%%%%%%%%%%%%%%%%%%%%%%%%%%%%%%%%%%%%%%%%%%%%%%%%%%%%%%%%%%%%
\subsection{Further results}\label{Sec:Further}

The analogue of the Lieb-Thirring conjecture does not hold in the context of Theorem~\ref{Thm:Main}, {\sl i.e.\/} for potentials such that $\lim_{i\to\infty}\lambda_i(V)=+\infty$.
%---------------------------------------------------------------------
\begin{prop}\label{Prop:LTconjecture} With the notations of Sections \ref{Sec:Intro} and \ref{Sec:DualCase}, for any $d\in\N^*$ and $\gamma>d/2$, 
\[
\CLTcd<\CLTd\;.
\]
Moreover, if $F$ and $G$ satisfy the assumptions of Theorem~\ref{Thm:Main-Gen}, then
\[
n\mapsto\sup_V\frac{\sum_{1\leq i\leq n}F(\lambda_i(V))}{\int_{\R^d}G(V(x))\,dx}=:{\mathcal C}^{(n)}(\gamma)
\]
forms a strictly increasing sequence.
\end{prop}
%---------------------------------------------------------------------
\proof The infimum $\CLTcd$ is achieved by a function $u_*$ with support in a ball $B(0,R)$ for some $R>0$, and a potential $V_*=c\,u_*^{4/(d-2(\gamma+1))}$ in $B(0,R)$ for some constant $c>0$, and $V_*=+\infty$ outside. The sequence of eigenvalues of $-\Delta+V_*$ is therefore given by the one of $-\Delta+V_*$ in $B(0,R)$ with zero Dirichlet boundary conditions on $\partial B(0,R)$. It is then straightforward to realize that 
\[
\sum_{i\in\N^*}\left[\lambda_i(V_*)\right]^{-\gamma}>\left[\lambda_1(V_*)\right]^{-\gamma}= \CLTcd\;.
\]
The general case follows from similar reasons.
\finprf

%%%%%%%%%%%%%%%%%%%%%%%%%%%%%%%%%%%%%%%%%%%%%%%%%%%%%%%%%%%%%%%%%%%%%%
%%%%%%%%%%%%%%%%%%%%%%%%%%%%%%%%%%%%%%%%%%%%%%%%%%%%%%%%%%%%%%%%%%%%%%
\section{Interpolation inequalities}\label{Sec:Gagliardo-Nirenberg-Systems}

Assume that $V$ is a potential on $\R^d$ such that the operator $-\Delta +V$ has an infinite sequence $\seq{\lambda_i(V)}$ of eigenvalues. Let $F$ and $G$ be two functions such that the inequality
\be{Ineq:Gen}
\sum_{i\in\N^*}F(\lambda_i(V))=\tr \left[ F\left(-\Delta +V\right)\right]\leq\int_{\R^d}G(V(x))\,dx
\ee
holds (see for instance Theorem~\ref{Thm:Main-Gen} for sufficient conditions). Let $\bar\lambda:=\lim_{i\to\infty}\lambda_i(V)$ and assume that 
\[
{\rm Spectrum}(-\Delta +V)\cap(-\infty,\bar\lambda)=\{\lambda_i(V)\;:\; i\in\N^*\}\;.
\]
Note that this includes the standard case of Lieb-Thirring inequalities, which corresponds to $\bar\lambda=0$ when $V$ is such that $-\Delta+V$ has infinitely many eigenvalues, and the case considered in Theorems~\ref{Thm:Main}~and~\ref{Thm:Main-Gen}: $\bar\lambda=+\infty$. 

Define $\sigma(s):=-F'(s)$ and $\beta(s):=-\int_0^s\sigma^{-1}(t)\,dt$. We may notice that 
\[
F(s)=\int_s^{\bar\lambda}\sigma(t)\;dt=\int_s^{\bar\lambda}(\beta')^{-1}(-t)\;dt\;.
\]
We assume that $F$ is convex on $(-\infty,\bar\lambda)$ and $C^1$ on $(-\infty,\bar\lambda)$ whenever it takes finite values. This implies  that $\beta$ is $C^1$, convex and we get
\[
F(s)=-\min_{\nu>0}\left[\beta(\nu)+\nu\,s\right]\;.
\]
Note indeed that, at a formal level,
\[
\frac d{ds}\left(\left[\beta(\nu)+\nu\,s\right]_{|\nu=(\beta')^{-1}(-s)}\right)=(\beta')^{-1}(-s)=\sigma(s)\;.
\]
Inequality \eqn{Ineq:Gen} can therefore be rewritten as 
\[
\sum_{i\in\N^*}\nu_i\ix{\left(|\nabla\psi_i|^2+V\,|\psi_i|^2\right)} + \sum_{i\in\N^*}\beta(\nu_i)+\ix{G(V(x))}\geq 0
\]
for any sequence of nonnegative occupation numbers $\seq{\nu_i}$ and any sequence $\seq{\psi_i}$ of orthonormal $L^2(\R^d)$ functions.

Let us proceed as in Section \ref{Sec:LT-GN} and optimize on $V$ for fixed $\bnu=\seq{\nu_i}$, $\bpsi=\seq{\psi_i}$. Assume further that $G'$ is invertible. Let 
\[
K[\bnu,\bpsi]:=\sum_{i\in\N^*}\nu_i\,|\nabla\psi_i|^2\quad\mbox{and}\quad \rho:=\sum_{i\in\N^*}\nu_i\,|\psi_i|^2\;,
\]
and define
\[
H(s):=-\left[G\circ(G')^{-1}(-s)+s\,(G')^{-1}(-s)\right]\,.
\]
It is straightforward to check as above that
\[
\frac{dH}{ds}(s)=-(G')^{-1}(-s)\;,
\]
and write
\[
H(s)=\int_s^0(G')^{-1}(-t)\,dt
\]
provided $(G')^{-1}$ is integrable on a neighborhood of $s=0_+$. 

The optimal potential $V$ has to satisfy
\[
G'(V)+\rho=0\;,
\]
so that
\[
\sum_{i\in\N^*}\nu_i\ix{V\,|\psi_i|^2}+\ix{G(V(x))}=-\ix{H(\rho(x))}
\]
Summarizing our computations, we have proved that \eqn{Ineq:Gen} can be rephrased as 
%-----------------------------------------------------------------------------
\begin{thm}\label{Thm:Interpolation} Under the above notations and assumptions, the following inequality holds:
\be{Ineq:InterpolationGen}
K[\bnu,\bpsi]+\sum_{i\in\N^*}\beta(\nu_i)\geq \ix{H(\rho)}
\ee
with $\rho=\sum_{i\in\N^*}\nu_i\,|\psi_i|^2$, where $\seq{\nu_i}$ is any nonnegative sequence of occupation numbers and $\seq{\psi_i}$ is any sequence of orthonormal $L^2(\R^d)$ functions.\end{thm}
%-----------------------------------------------------------------------------
Written with such a generality, the result is maybe not as striking as when it applies to the various examples of Section \ref{Sec:Stability}, for which all the assumptions made above can be verified. To keep the generality of our result, we will not try to give sufficient conditions on $\beta$ and $V$ for which all these assumptions can be established and prefer to state three applications corresponding for the function $\beta$ to $\beta(\nu)=Const\cdot\nu^m$ with $m\in(-\infty,0)\cup(1,+\infty)$, $\beta(\nu)=-Const\cdot\nu^m$ with $m\in (0,1)$ and $\beta(\nu)=\nu\log\nu-\nu$.

\par\medskip\noindent\underline{{\sl Example 1.\/}} Let $m>1$, which corresponds to the setting of the standard Lieb-Thirring inequality~\eqn{Ineq:LT}, and consider $\beta(\nu):=c_m\,\nu^m$, $c_m:=(m-1)^{m-1}m^{-m}$, $m=\frac\gamma{\gamma-1}$, $F(s)=(-s)^\gamma$ and $G(s)=\CLT(-s)^{\gamma+d/2}$. Define
\[
q:=\frac{2\gamma+d}{2\gamma+d-2}\quad\mbox{and}\quad {\mathcal K}^{-1}:=q\left[\CLT\left(\gamma+\frac d2\right)\right]^{q-1}.
\]

%-----------------------------------------------------------------------------
\begin{corollary}\label{Cor:Interp1} With the above notations, for any $m\in(1,+\infty)$, the following optimal inequality holds:
\[
K[\bnu,\bpsi]+c_m\sum_{i\in\N^*}\nu_i^m\geq{\mathcal K}\ix{\rho^q}\;.
\]
\end{corollary}
%-----------------------------------------------------------------------------
Using the scaling invariance, we can reformulate this result as follows. If we replace $\psi_i(x)$ by $\lambda^{-d/2}\psi_i(x/\lambda)$ and $\nu_i$ by $\lambda^{d(1-1/q)}\nu_i$, the right hand side of the above inequality is invariant. An optimization of the left hand side shows that
\[
\Bigg(K[\bnu,\bpsi]\Bigg)^\theta\left(\sum_{i\in\N^*}\nu_i^m\right)^{(1-\theta)}\geq{\mathcal L}\ix{\rho^q}\;,
\]
where $\theta=\frac d{2(\gamma-1)+d}$ and ${\mathcal L}$ can be explicitly computed in terms of ${\mathcal K}$, $d$ and $\gamma$.

The case $m=\frac\gamma{\gamma-1}\in(-\infty,0)$, which corresponds to $\gamma\in (0,1)$ and $\beta(\nu):=c_m\,\nu^m$, $c_m:=-(1-m)^{m-1}|m|^{-m}$ is formally covered with the same constants, but does not enter in our framework for infinite systems (see Example 1, Section 3.4). Notice that $q$ varies in the range $(1,1+\frac d2)$ for $m>1$ and $(1+\frac d2,\frac d{d-2})$ if $m<0$. The case $\gamma=1$, $q=1+\frac d2$ is not covered.

\par\medskip\noindent\underline{{\sl Example 2.\/}} If $m\in (0,1)$, which corresponds to the setting of Theorem~\ref{Thm:Main}, and $\beta(\nu):=-c_m\,\nu^m$, $c_m:=(1-m)^{m-1}m^{-m}$, $m=\frac\gamma{\gamma+1}$, $F(\lambda)=\lambda^{-\gamma}$ and $G(s)=\CMain\,s^{d/2-\gamma}$. Define
\[
q:=\frac{2\gamma-d}{2(\gamma+1)-d}\in(0,1)\quad\mbox{and}\quad {\mathcal K}^{-1}:=q\left[\CMain\left(\gamma-\frac d2\right)\right]^{q-1}\;.
\]
Notice that due to the restriction $\gamma>d/2$, the range of $m$ is reduced to the interval $(\frac d{d+2},1)$.
%-----------------------------------------------------------------------------
\begin{corollary}\label{Cor:Interp2} With the above notations, for any $m\in(\frac d{d+2},1)$, the following optimal inequality holds:
\[
K[\bnu,\bpsi]+{\mathcal K}\ix{\rho^q}\geq c_m\sum_{i\in\N^*}\nu_i^m\;.
\]
\end{corollary}
%-----------------------------------------------------------------------------
Using the scaling invariance, we can also reformulate this result as follows. If we replace $\psi_i(x)$ by $\lambda^{-d/2}\psi_i(x/\lambda)$ but dont change $\nu_i$, the right hand side of the above inequality is of course invariant. An optimization of the left hand side shows that
\[
\Bigg(K[\bnu,\bpsi]\Bigg)^\theta\left(\ix{\rho^q}\right)^{(1-\theta)}\geq{\mathcal L}\sum_{i\in\N^*}\nu_i^m\;,
\]
where $\theta=\frac d{2(\gamma+1)}$ and ${\mathcal L}$ can be explicitly computed in terms of ${\mathcal K}$, $d$ and $\gamma$.

\par\medskip\noindent\underline{{\sl Example 3.\/}} If $\beta(\nu):=\nu\log\nu-\nu$, then $\beta'(\nu)=\log\nu=-\lambda$, $F(s)=e^{-s}$ and $G(s)=(4\pi)^{-d/2}\,e^{-s}$. Inequality \eqn{Ineq:InterpolationGen} is a logarithmic Sobolev inequality for systems:
%-----------------------------------------------------------------------------
\begin{corollary}\label{Cor:InterpLogSob} With the above notations, the following optimal inequality holds:
\[
K[\bnu,\bpsi]+\sum_{i\in\N^*}\nu_i\log\nu_i\geq \ix{\rho\log\rho}+\frac d2\,\log(4\pi)\,\ix\rho\;.
\]
\end{corollary}
%-----------------------------------------------------------------------------
As above, an optimization under a scaling preserving the $L^2$ norm of $\psi$ and leaving $\nu_i$ invariant allows to write
\[
\ix{\rho\log\rho}\leq \sum_{i\in\N^*}\nu_i\log\nu_i+\frac d 2\,\log\left(\frac e{2\pi\,d}\,\frac{K[\bnu,\bpsi]}{\ix{\rho}}\right)\ix{\rho}\;.
\]

\medskip Note that we immediately recover the Gagliardo-Nirenberg inequalities of Section~\ref{Sec:Stability} by taking $\nu_1=1$, $\nu_i=0$ for any $i\geq 2$, in case of Examples 1 and 2, but with {\sl a priori\/} non optimal constants, at least in the case of Example 2. The proof of Proposition~\ref{Prop:Interpolation} follows for the same reason.

%---------------------------------------------------------------------
\bigskip\noindent{\bf Remark. }{\sl We notice that the limit case $\gamma=0$ for $d\geq 3$ is not covered, even as a limit case. For $\nu_i=1$, for $i=1$, $2$,\ldots $N$, and $\nu_i=0$ otherwise, {\sl a Sobolev type inequality for orthonormal functions\/} has been given in \cite{MR1112575,MR1037324} in the case which corresponds to the critical Sobolev embedding $H^1(\R^d)\hookrightarrow L^{2d/(d-2)}(\R^d)$. By taking the occupation numbers $\nu_i$ into account, we always achieve optimal inequalities which are related in a natural way to some corresponding optimal Lieb-Thirring inequalities as long as $\gamma$ is positive. To a large extend, this improves the known results for orthonormal and sub-orthonormal systems \cite{MR1112575,MR920485,MR1135275}}.
%---------------------------------------------------------------------

%%%%%%%%%%%%%%%%%%%%%%%%%%%%%%%%%%%%%%%%%%%%%%%%%%%%%%%%%%%%%%%%%%%%%%
%%%%%%%%%%%%%%%%%%%%%%%%%%%%%%%%%%%%%%%%%%%%%%%%%%%%%%%%%%%%%%%%%%%%%%
\bigskip\bigskip\noindent{\sl Acknowledgments.\/} {\small J.D. and E.P. are partially supported by ECOS-Conicyt under grants no. C02E06 and no. C02E08 and by European Programs HPRN-CT \# 2002-00277 \& 00282. M.L. is partially supported by U.S. National Science Foundation grant DMS 03-00349. P.F. is partially supported by ECOS-Conicyt under grant no. C02E08, Fondecyt Grant 1030929 and FONDAP de Matem\'aticas Aplicadas.}

\bigskip\noindent{\sl \copyright~2005 by the authors. This paper may be reproduced, in its entirety, for non-commercial purposes.}
%%%%%%%%%%%%%%%%%%%%%%%%%%%%%%%%%%%%%%%%%%%%%%%%%%%%%%%%%%%%%%%%%%%%%%
%%%%%%%%%%%%%%%%%%%%%%%%%%%%%%%%%%%%%%%%%%%%%%%%%%%%%%%%%%%%%%%%%%%%%%
%%%%%%%%%%%%%%%%%%%%%%%%%%%%%%%%%%%%%%%%%%%%%%%%%%%%%%%%%%%%%%%%%%%%%%
%\nocite*
%\bibliographystyle{siam} 
%\bibliography{References.bib}
%\end{document}
%%%%%%%%%%%%%%%%%%%%%%%%%%%%%%%%%%%%%%%%%%%%%%%%%%%%%%%%%%%%%%%%%%%%%%
%%%%%%%%%%%%%%%%%%%%%%%%%%%%%%%%%%%%%%%%%%%%%%%%%%%%%%%%%%%%%%%%%%%%%%

\end{document}